\documentclass[copyright]{eptcs}
\providecommand{\event}{AMCA-POP 2010} 

\providecommand{\firstpage}{1}

\usepackage{graphicx}

\usepackage{amssymb}
\usepackage{amsthm}

\usepackage{amsmath}
\usepackage{latexsym}
\usepackage[mathscr]{euscript}
\usepackage{color}

\newcommand{\cS}{{\mbox{$\mathscr{S}$}}}

\newcommand{\cX}{{\mbox{$\mathcal{X}$}}}

\newcommand{\Nat}{\mathbb{N}}
\newcommand{\Natpos}{\mathbb{N}^{> 0}}
\newcommand{\Integer}{\mathbb{Z}}

\newcommand{\inv}{\ensuremath{\mathit{Inv}}}

\newtheorem{theorem}{Theorem}[section]
\newtheorem{definition}[theorem]{Definition}

\newtheorem{proposition}[theorem]{Proposition}

\newcommand{\range}{\mathtt{range}}
\newcommand{\steps}{\mathtt{Steps}}

\title{An Individual-based Probabilistic Model for Fish Stock Simulation}
\author{Federico Buti \qquad
\institute{School of Science and Technology\\ University of Camerino, Italy}
\email{federico.buti@unicam.it} \and
Flavio Corradini \qquad
\institute{School of Science and Technology\\ University of Camerino, Italy}
\email{flavio.corradini@unicam.it} \and
Emanuela Merelli \qquad
\institute{School of Science and Technology\\ University of Camerino, Italy}
\email{emanuela.merelli@unicam.it} \and
Elio Paschini \qquad
\institute{CNR - Institute of Marine Sciences\\ Ancona, Italy}
\email{e.paschini@ismar.cnr.it} \and
Pierluigi Penna \qquad
\institute{CNR - Institute of Marine Sciences\\ Ancona, Italy}
\email{p.penna@an.ismar.cnr.it} \and
Luca Tesei
\institute{School of Science and Technology\\ University of Camerino, Italy}
\email{luca.tesei@unicam.it}}

\def\titlerunning{An Individual-based Probabilistic Model for Fish Stock Simulation}
\def\authorrunning{F. Buti et al.}
\begin{document}
\maketitle

\begin{abstract}
We define an individual-based probabilistic model of a sole (Solea solea) behaviour. The individual model is given in terms of an Extended Probabilistic Discrete Timed Automaton (EPDTA), a new formalism that is introduced in the paper and that is shown to be interpretable as a Markov decision process. A given EPDTA model can be probabilistically model-checked by giving a suitable translation into syntax accepted by existing model-checkers. In order to simulate the dynamics of a given population of soles in different environmental scenarios, an agent-based simulation environment is defined in which each agent implements the behaviour of the given EPDTA model. By varying the probabilities and the characteristic functions embedded in the EPDTA model it is possible to represent different scenarios and to tune the model itself by comparing the results of the simulations with real data about the sole stock in the North Adriatic sea, available from the recent project SoleMon. The simulator is presented and made available for its adaptation to other species.
\end{abstract}

\section{Introduction}

Ecosystems are composed of living animals, plants and non-living structures that exist together and interact with each other. Fish are part of the marine ecosystem and interact closely with their physical, chemical and biological environment. They are inter-dependent with the ecosystem that provides the right conditions for their growth, reproduction and survival. Conversely, they are a source of food for other animals and form an integral part of the marine food web.

The fishing activity impacts both on the fish stocks and on the ecosystem within which they live. The Ecosystem Approach to Fisheries (EAF) \cite{Garcia2003} recognises that fisheries have to be managed as part of their ecosystem and that the impact on the environment should be limited as much as possible. Part of this approach is the fish stock assessment. A ``stock'' is a population of a species living in a defined geographical area with similar biological parameters (e.g.\ growth, size at maturity, fecundity etc.) and a shared mortality rate. Its aim is to provide information to managers on the state and life history of the stocks. This information is used into the decision making process. Stock assessment can be made using mathematical and statistical models to examine the history of the stock and to make quantitative predictions in order to address the following questions: 1) What is the current state of the stock? 2) What has happened to the stock in the past? 3) What will happen to the stock in the future if different management choices are made? Fisheries employs a wide variety of recognised assessment models and statistical methods to assess the stocks of fish. If we know about the stock size (biomass) and the biology of the fish stock, we can estimate how many fish can be safely removed from the stock in order to ensure a sustainable resource.

Using the data from a recent research project \cite{Solemon,Fabi2007} on the common sole (\textit{Solea solea}), it was possible to obtain new information on the biology of this fish. These data are the input of mathematical models based on equations determining the stock assessment of this species. This makes possible to regulate the fishing effort in order to avoid overfishing. In this work we want to introduce a somewhat new way of addressing fish dynamic population modelling and fish stock assessment. The main characteristic of our approach is that it is individual-based, that is to say, every single individual of the population under study is considered as an independent entity and the dynamics of the overall population living in a given environment \textit{emerges} from the individual interactions and behaviours. Every aspect of the population can then be observed and measured in a simulation environment. This last aspect permits the tuning and the validation of the individual model by using existent experimental data.

Since systems biology was proposed as a challenge of a new way of understanding biology, it has involved biologists, physicians, mathematicians, physicists, computer scientists and engineers. In particular, in the computer science community a lot of models, languages, approaches and methodologies have been applied in a biological context, and several formalisms have been specifically developed for describing different aspects of biological systems.
In \cite{Barbuti2009}, authors extend P systems with features typical of timed automata with the aim of describing periodic environmental events such as seasons or periodical hunts/harvests. In \cite{512} it is proposed a modelling framework based on P systems and it is applied to the modelling of the dynamics of some scavenger birds in the Pyrenees. This model considers information about the feeding of the population. In \cite{Barbuti2010}, a spatial extension of P systems is introduced and an example of the evolution of ‘‘ring species’’, based on small changes between geographically contiguous populations, is modelled. Authors of \cite{1428721} present a process algebraic approach to the modelling of population dynamics. Currently no time characterisation can be provided of the modelled biological environment because the calculus has not a notion of time. Stamatopoulou et al.\  provide, in \cite{Stramato2} and \cite{Stamatopoulou_operas:a}, models based on X-machines and P systems for biological-inspired systems such as colony of ants or bees, flocks of birds and so on. Besozzi et al.\ \cite{Besozzi} model metapopulations (which are ecological models describing the interactions and the behaviour of populations living in fragmented habitats) by means of dynamical probabilistic P systems where additional structural features have been defined (e.g., a weighted graph associated with the membrane structure and the reduction of maximal parallelism). Such a work effectively uses many regions to model an ecological system, thus it really exploits the advantages of the membrane structure. In \cite{Gheorghe01computationalmodels} authors model the behaviour of a bee colony as a society of communicating agents acting in parallel and synchronising their behaviour. Two models are provided: one is based on P systems while the other is based on X-machines but no tool, thus no actual results, are available to compare the behaviour of the two models. Finally, in an older work of Bahr and Bekoff \cite{BAHR} authors model a flock in terms of cellular automata; although interesting, theirs work concentrates only on the \textit{vigilance} of the flock and how it is affected by internal and external factors (such as flock size, number of obstacles and so on).

The individual-based vision is quite natural in the computer science world, since notions such as process, component, activity, flow, interaction all can be easily related to an executor or a virtual entity. When adapting these notions to a biological scenario it is natural to reason in terms of entities that ``do something'' and probably collaborate to make the whole system function well. On the other side, biologists have often a view that abstracts from single entities preferring to reason in terms of aggregated variables, for which a continuous domain can be adopted and that are related by differential equations (ODE, PDE). Consequently, the available data from observations and experiments follow this way of thinking and are not directly interpretable in an individual-based setting. To bridge the gap there is the need to develop methodologies and software systems that make the two worlds interact and somehow work in synergy to transport the advantages of each view into the other and vice versa.

In this paper we define a formalism called Extended Probabilistic Discrete Timed Automata (EPDTA) that is a variant of probabilistic timed automata \cite{Kwiatkowska2002}. It simplifies the time domain, that is discrete, but introduces integer and boolean variables in the state. The formalism is shown to be interpretable as a Markov decision process and also easily translatable to a syntax that is accepted by the probabilistic model-checker PRISM \cite{Kwiatkowska2002a,Kwiatkowska2009}. A model of the behaviour of the common sole (\textit{Solea solea}) living in the North Adriatic sea is then given in terms of an EPDTA by using available data from a recent project \cite{Solemon,Fabi2007}. After the individual behaviour is defined we introduce a simulation environment that is agent-based and derives from the one developed in \cite{Penna2010}. Essentially, it creates a Multi-Agent System (MAS) in which every agent represents a sole whose internal (probabilistic) behaviour is given by the individual model. The MAS permits a precise monitoring of all the events occurring in the virtual square kilometre of sea that is simulated. The simulator, called FIShPASs (FIshing Stock Probabilistic Agent-based simulator), is available \cite{FishpassSite} and easily adaptable to simulate other species.

The paper is organised as follows: Section~\ref{sec:epdta} defines EPDTAs and gives their semantics as a Markov decision process. Section~\ref{sec-model} shows a particular EPDTA representing the individual behaviour of a sole living in the North Adriatic sea. Section~\ref{sec:simul} introduces the simulator as a MAS in which each agent implements the individual probabilistic behaviour described in Section~\ref{sec-model} and shows the preliminary simulation results that have to be tuned/validated using the real observation data available form the SoleMon project. Section~\ref{sec:concl} concludes, describing some future work.

\section{Extended Probabilistic Discrete Timed Automata}
\label{sec:epdta}

\begin{figure}[t]
  \begin{center}
    \includegraphics[height=4cm]{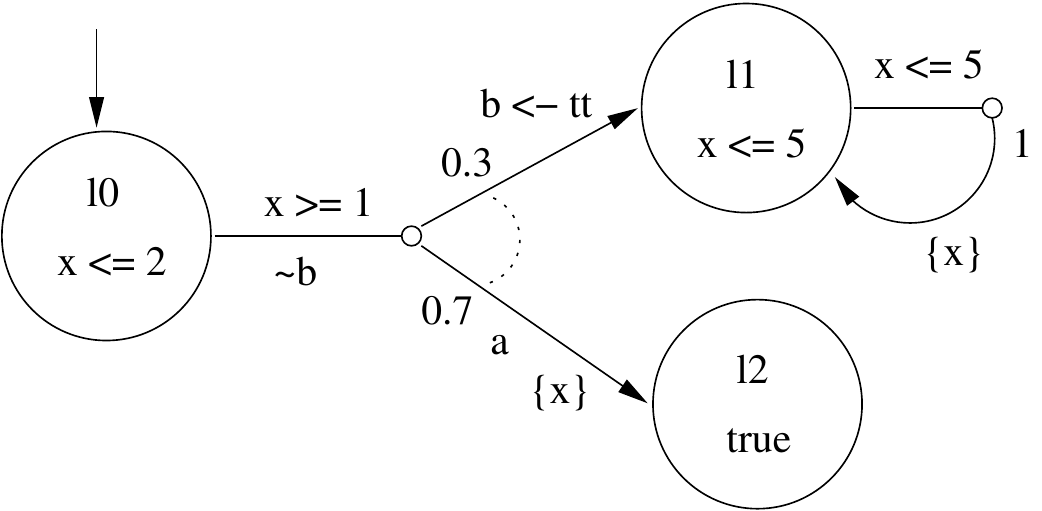}
  \end{center}
\caption{A simple EPDTA.}
\label{fig-epdta1}
\end{figure}

In this section we introduce EPDTAs, a variant of probabilistic timed automata that we need for our purpose. We then show that an automaton of this kind can be interpreted as a Markov decision process and that can be translated easily into one handled by the model checker PRISM \cite{Kwiatkowska2002a,Kwiatkowska2009}.

Briefly, a timed automaton (TA) \cite{Alur1994} is an automaton equipped with real-valued clock variables and such that transitions are guarded by clock constraints. The control flows from a state to another instantaneously if and only if the guard of that transition is enabled. Each transition has an action associated and can reset some clocks. While the control stays in a state, time elapse i.e.\ clock values increase. Possible conditions in states, called invariants, can prevent the passage of time forcing the control to exit from that state with one of the enabled transitions.

Following the approach of \cite{Bianco1995}, probability has been introduced into timed automata yielding probabilistic timed automata (PTA) \cite{Beauquier2003,Kwiatkowska2002}. In this case every transition from a state has a clock constraint as a guard, but then the action, reset and destination state is given by a finite probability distribution. Thus, every step of a PTA consists in a resolution of non-determinism among different enabled transitions, passage of time included, followed by a probabilistic choice of the action, reset and destination according to the given distribution.

An EPDTA is essentially a PTA with the set $\Nat$ of natural numbers as time domain and in which the locations are enriched with a finite set of boolean and finite-range integer variables. Clocks, that are considered similar to integer variables with range $[0,\infty]$, grow at discrete steps of length 1. We also add a subset of actions that are called \emph{urgent} and that must be executed as soon as they are enabled. The motivation of these variants are essentially given by the peculiarities of the models of individuals in ecosystems: their state-changes are typically modelled in terms of transitions that happen at a time scale of years, months and, in the finer grain, weeks. Thus, there is no need of continuous time. Moreover, each individual has some characteristics, e.g. age, sex, length, weight, fertility, last time of reproduction, etc. that are easily representable by integer (with finite range) and boolean variables and that influence its behaviour. Last characteristic of this meta-model is a constant $\mathtt{MAX\_TIME} \in \Nat$ that represents the maximum number of time steps that the automaton can perform. This means that each clock has, actually, a range $[0,\mathtt{MAX\_TIME}]$. Since this constant can be chosen arbitrarily large, this requirement does not limit the generality of the meta-model both if it is used in a simulation environment and if it is used for model checking. On the other hand it permits to give a finite range to all variables of the model, clocks included.

Now we define in detail the syntax and the semantics of EPDTAs. This part is quite technical and can be skipped by the non-familiar reader. Section~\ref{sec-model} will present the model of the Solea solea behaviour as a particular EPDTA that can be quite intuitively understood even without knowing all the technical details. The idea of clock variables is central in the framework of timed automata and it is imported in our meta-model. A {\em clock} is a variable that takes values from the set $\Nat$. Clocks measure time as it elapses, all clocks of a given system advance at the same pace and clock variables are ranged over by $x, y, z, \ldots$ We use $X, X', \ldots$ to denote finite sets of clocks. A {\em clock valuation} over $X$ is a function assigning a natural number to every clock. The set of valuations of $X$, denoted by $V_X$, is the set of total functions from $X$ to $\Nat$. Clock valuations are ranged over by $\nu, \nu', \ldots$. Given $\nu \in V_X$ and $n \in \Nat$, we use $\nu + n$ to denote the valuation that maps each clock $x \in X$ into $\nu(x)+n$.

Clock variables, like other variables, can be assigned during the evolution of the system when certain actions are performed. The assignment consists in instantaneously set the value of a variable to a new value. Clock variables are always assigned to 0, i.e.\ they are reset. Immediately after this operation a clock restarts to measure time at the same pace as the others. The reset is useful to measure the time elapsed since the last action/event that reset the
clock. Given a set $X$ of clocks, a {\em reset} $\gamma$ is a subset of $X$. The set of all resets of clocks in $X$ is denoted by $\Gamma_X$ and reset sets are ranged over by $\gamma, \gamma', \ldots$ Given a valuation $\nu \in V_X$ and a reset $\gamma$, we let $\nu \backslash \gamma$ be the valuation that assign the value 0 to every clock in $\gamma$ and assign $\nu(x)$ to every clock $x \in X \backslash \gamma$.

We need also to consider a finite set $B$ of boolean variables, ranged over by $b, b', \ldots$, a finite set $I$ of integer variables, ranged over by $v, v', \ldots$, together with a range assignment function $\range:I \leftarrow \Integer \times \Integer$ such that if $\range(v) = (z_1,z_2)$ then $z_1 \leq z_2$. Finally, we need a finite set $F$ of totally specified functions, ranged over by $f, f', \ldots$, that we use as tables in which constants values are collected and where then they can be retrieved by applying each function to values in its domain (essentially they are tables of probability values or array of constants). Such tables can contain rational numbers. If they are involved in integer operations they are rounded to the closest integer.

The grammars introduced in the following define the syntax of a first-order language in which very usual functions and relations are present. The language can express boolean and arithmetic expressions. Moreover, we define a syntax for expressing assignments to variable of corresponding type, clock constraints and guards. $Bexp \; ::= \; \mathit{tt} \mid \mathit{ff} \mid b \mid Bexp \wedge Bexp \mid \sim Bexp \mid Aexp = Aexp \mid Aexp <= Aexp \mid Aexp < Aexp \mid (Bexp)$, $Aexp \; ::= \; z \mid v \mid f(Aexp, Aexp, ...)\footnote{According to what said above, this can be considered a constant. Of course the arguments of the function must be of the right number and of the right type.} \mid Aexp + Aexp \mid Aexp * Aexp \mid Aexp - Aexp \mid Aexp / Aexp\footnote{Integer division.} \mid Aexp \% Aexp\footnote{Rest of integer division.} \mid (Aexp)$  where $z \in \Integer$. Assignments are of the form $Assign \; ::= \; b \leftarrow Bexp \mid v \leftarrow Aexp \mid Assign, Assign$. Boolean expressions are ranged over by $\beta, \beta', \ldots$, arithmetic expressions are ranged over by $\alpha, \alpha', \ldots$ and assignments are ranged over by $\eta, \eta', \ldots$.  The timed behaviour of the system is expressed using constraints on the actual values of the clocks. Given a set $X$ of clocks, the set $\Psi_X$ of {\em clock constraints} over $X$ is defined by the following grammar: $\psi \; ::= \; \mathit{true} \mid \mathit{false} \mid x \; \# c \mid x - y \; \# \; c \mid \psi \wedge \psi$ where $x, y \in \cX$, $c \in \Nat$, and $\# \in \{<,>,\leq, \geq, =\}$. Finally, guards, ranged over by $g, g', \ldots$ are defined as $Guard \; ::= \; \psi \mid Bexp \mid Guard \wedge Guard$. As usual, we use the name of the syntactic category to denote the set of the generated objects. Thus, for instance, $Guard$ represents the set of all strings that are well-formed guards.

Given sets $B$, $I$, we define an interpretation $\iota$ as a function assigning a value to every variable in $B$ and $I$. By means of an interpretation $\iota$ we can evaluate a boolean expression $\beta$ or an arithmetic expression $\alpha$ in the standard sense; we denote with ${\cal E}_\iota(\beta)$ the boolean value of $\beta$ and with ${\cal E}_\iota(\alpha)$ the integer value of $\alpha$ both using the interpretation $\iota$. Moreover, we can define a satisfaction relation $\models$ such that $\nu \models \psi$ if the values of the clocks in $\nu$ satisfy the constraint $\psi$ in the natural interpretation. Finally, the satisfaction relation can be extended naturally on guards: $\iota, \nu \models g$.

An assignment $\eta$ is evaluated as a change in the interpretation $\iota$. We denote with ${\cal A}(\iota,\eta)$ a new interpretation $\iota'$ in which the variables that are assigned in $\eta$ are all\footnote{Note that we suppose that $\eta$ does not contain more than one assignment for each variable.} changed with the corresponding values, evaluated from $\iota$ in the above sense.

Given a set $H$ let us denote by $\mu(H)$ the set of \emph{finite} probability distributions over $H$ i.e.\ $\mu(H)$ contains functions $p \colon H \rightarrow [0,1]$ such that $\sum_{h \in H}p(h) = 1$ and the set $\{h \in H \mid p(h) > 0 \}$ is finite. A probability distribution $p$ can be represented as follows: $p = [h_1 \mapsto p_1, \ldots, h_n \mapsto p_n]$ where the $h_i$'s are exactly all elements of $H$ that have $p(h_i) = p_i > 0$.

\begin{definition}[EPDTA]
An \emph{extended probabilistic discrete timed automaton} $T$ is a tuple\\ $(Q,\Sigma,B,I,X,\mathbb{E}, \mathbb{U}, q_0, \iota_0, \inv)$, where: $Q$ is a finite set of locations, $\Sigma$ is a finite alphabet of actions, $B$ is a finite set of boolean variables, $I$ is a finite set of finite-range integer variables, $X$ is a finite set of clocks, $\mathbb{E}$ is a finite set of edges, $\mathbb{U}$ is a finite set of \emph{urgent} edges, $q_0$ is the initial location, $\iota_0$ is the initial interpretation of the variables of $B \cup I$, $\mathtt{MAX\_TIME}$ is the maximum time of evolution and $\inv$ is a function assigning to every $q \in Q$ an \emph{invariant}, i.e.\ a clock constraint $\psi$ such that for each clock valuation $\nu \in V_X$ and for each $n \in \Natpos$, $\nu + n \models \psi \Rightarrow \nu \models \psi$. Constraints having this property are called \emph{past-closed}.

Each edge $e \in \mathbb{E} \cup \mathbb{U}$ is a tuple in $Q \times Guard \times \mu(\Sigma \times Assign \times \wp(X) \times Q)$. If $e=(q,g,prob)$ is an edge, $q$ is the {\em source}, $g$ is the {\em guard} and $prob$ is the {\em distribution}. If  $prob((a,\eta,\gamma,q')) > 0$ then there is a possibility for the automaton to reach the \emph{target} location $q'$ performing the \emph{action} $a$, the {\em assignment} $\eta$ and the \emph{reset} $\gamma$.
\end{definition}

\begin{figure}[t]
  \begin{center}
    \includegraphics[width=13.5cm]{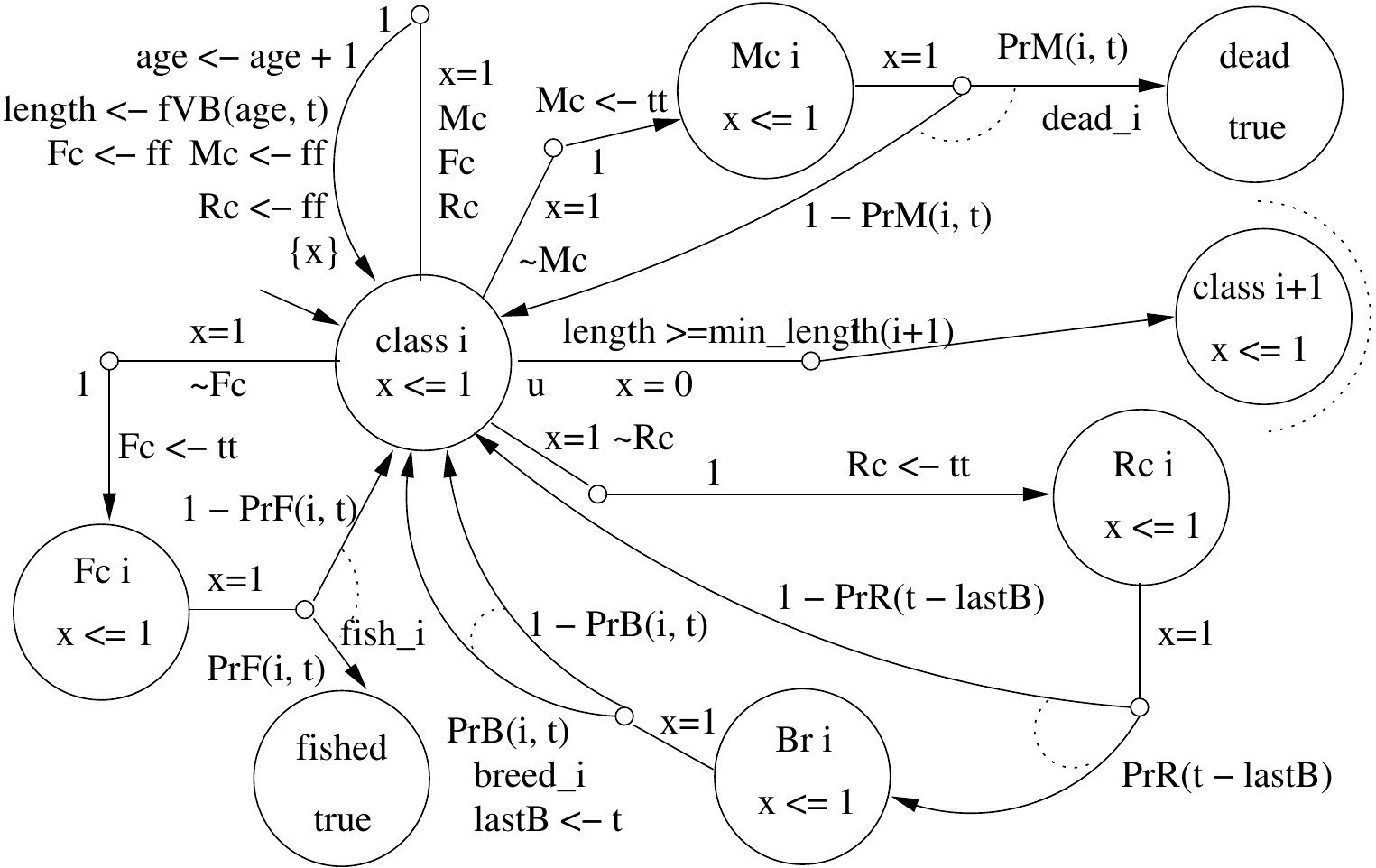}
  \end{center}
\caption{EPDTA representing the behaviour of the sole when it is in class $i$. The other classes are equal.}
\label{fig-solea}
\end{figure}

Figure~\ref{fig-epdta1} shows an EPDTA with three locations $l0,l1,l2$. The set of clocks is $\{x\}$, the alphabet is $\{a\}$, $l0$ is the initial state, and the invariant of state $l0$ is $x \leq 2$. There is an edge starting from location $l0$ with a guard that is the conjunction of the clock constraint $x \geq 1$ and the boolean expression $\sim b$, where $b \in B$. At the edge it is associated a distribution $[(l1,a,\epsilon,\{x\}) \mapsto 0.7, (l2,\epsilon, b \leftarrow tt, \{ \}) \mapsto 0.3]$, where $\epsilon$ is the empty string. From $l1$ there is an edge in which the probability distribution is trivial. This transition is equivalent to a ``classical'' one.

The semantics of an EPDTA is a Markov decision process. A Markov decision process (MDP) is a pair $(\cS, \steps)$ where $\cS$ is a set of states and $\steps$ is a function giving for each state $s$ a set of probability distributions. Each $p \in \steps(s)$ is a discrete probability distribution in $\mu(\cS)$, saying the probability of each state of being the next state $s'$ in the process. A given MDP evolves as follows: at each step it is in a state $s$. Firstly it performs a non-deterministic choice to decide which distribution $p \in \steps(s)$ it will apply. Then it performs a probabilistic choice to go to a new state $s'$ according to the chosen $p$. The process then cycles again.

The semantics of $T = (Q,\Sigma,B,I,X,\mathbb{E}, \mathbb{U}, q_0, \iota_0, \inv)$ is a MDP $(S, \steps)$ where the set of states $S$ is the set of all the tuples of the form $(q,\nu,\iota)$ where $q \in Q \cup \{stop\}$\footnote{The fresh location $stop$ is added to terminate the activity of the automaton when the maximum time is reached.}, $\nu \in V_{X \cup \{t\}}$ is a valuation of the set of clocks $X$ augmented with a fresh clock $t$ that is never reset\footnote{This clock is needed to count the global elapsed time.} and $\iota$ is an interpretation of the variables in $B \cup I$. Note that if we fix a $\mathtt{MAX\_TIME}$ as the maximum time step for the system evolution then the set of states if finite as $Q$ is finite and all variables, clock included, have a finite range.

For every state $s=(q,\nu,\iota)$ the set of distributions $\steps(s)$ is determined by the following rules:
\begin{description}
  \item[Stop] \textbf{if} $\nu(t) = \mathtt{MAX\_TIME}$ \textbf{then} $[(stop,\nu,\iota) \mapsto 1] \in \steps(s)$
  \item[Time] \textbf{if} $\nu + 1 \models \inv(q)$ \textbf{and} $\nu(t) + 1 \leq \mathtt{MAX\_TIME}$ \textbf{and} $(\forall (q',g',prob') \in \mathbb{U} (q'=q \Rightarrow \iota, \nu \not \models g'))$ \textbf{then} $[(q,\nu+1,\iota) \mapsto 1] \in \steps(s)$
  \item[Urgent] \textbf{if} $\nu(t) \neq \mathtt{MAX\_TIME}$ \textbf{and} $(q,g,prob)\in \mathbb{U}$ \textbf{and} $\iota, \nu \models g$ \textbf{then} $[(q_1',\nu \backslash \gamma_1, {\cal A}(\iota,\eta_1)) \mapsto p_1, \ldots,\\ (q_n',\nu \backslash \gamma_n, {\cal A}(\iota,\eta_n)) \mapsto p_n] \in \steps(s)$ \textbf{where} $prob = [(a_1,\eta_1, \gamma_1, q'_1) \mapsto p_1, \ldots, (a_n,\eta_n, \gamma_n, q'_n) \mapsto p_n]$
  \item[Non-Urgent] \textbf{if} $\nu(t) \neq \mathtt{MAX\_TIME}$ \textbf{and} $(q,g,prob)\in \mathbb{E}$ \textbf{and} $\iota, \nu \models g$ \textbf{and} $(\forall (q',g',prob') \in \mathbb{U} (q'=q \Rightarrow \iota, \nu \not \models g'))$ \textbf{then} $[(q_1',\nu \backslash \gamma_1, {\cal A}(\iota,\eta_1)) \mapsto p_1, \ldots, (q_n',\nu \backslash \gamma_n, {\cal A}(\iota,\eta_n)) \mapsto p_n] \in \steps(s)$ \textbf{where} $prob = [(a_1,\eta_1, \gamma_1, q'_1) \mapsto p_1, \ldots, (a_n,\eta_n, \gamma_n, q'_n) \mapsto p_n]$
\end{description}

When rule \textbf{Stop} is applicable then no other rule is applicable. The process goes unconditionally to the $stop$ location in which it stops. Rule \textbf{Time} lets one time unit to elapse, provided that the invariant of the current location will be satisfied at the reached state, that the maximum time was not reached and that no urgent transitions (i.e.\ that do not permit time passing) are enabled. Rule \textbf{Urgent} inserts in $\steps(s)$ all possible distributions that derive from urgent transitions. Note that in case of more than one urgent transitions enabled all are inserted in $\steps(s)$ and thus a non-deterministic choice is done among them by the MDP. The resulting distributions are essentially the same of the original automaton, but here all the operations are performed on the clocks and on the variables to calculate the resulting state. The last rule \textbf{Non-Urgent} is applicable only if there are not urgent transitions enabled. The effect is the same of the urgent case and among different enabled non-urgent transitions a non-deterministic choice is done by the MDP.

\begin{proposition}
  Given an EPDTA $T$ and a natural number $\mathtt{MAX\_TIME}$ it is possible to construct a Markov decision process $\Pi$ in the syntax readable by the model checker PRISM such that $\Pi$ and the semantics of $T$ are the same Markov decision process.
\end{proposition}

We plan to provide an automatic tool for this translation inside our simulator FIShPASs (see Section~\ref{sec-fishpass}). This is very important because having the PRISM equivalent model improves the tests that the biologists can do against the probabilities put in the model itself. This is because quantitative questions can be asked to the model checker to test hypothesis made about the model or to validate it with available real data. A very powerful and useful logic language, Probabilistic CTL (PCTL) \cite{Kwiatkowska2002a,Kwiatkowska2008}, is suitable for expressing such questions.

\section{Sole Characteristics and Behaviour: an EPDTA Model}
\label{sec-model}

The body of the common sole (\textit{Solea solea}) is egg-shaped and flat \cite{FAOAdriamed,Fabi2007,Bolognini2008}. The maximum body height is equal to 1/3 of the total length. The eyes are on the right side, the upper one slightly anterior to the lower. Both pectoral fins are well developed, the left one being somewhat smaller than the right one. The dorsal fin begins anterior to the eyes, by the mouth. The last rods of the dorsal and the anal fins are connected to the caudal fin, which is round. The colour on the eyed side of the body is greyish-brown to reddish-brown, with large and diffused dark spots. The pectoral fin has a blackfish spot at its distal half. The posterior margin of the caudal fin is generally dark. This common sole species lives in the eastern Atlantic, from Scandinavia to  Senegal and in the entire Mediterranean. It is rare in the Black Sea.

Here we present an individual model of a sole living in the North Adriatic sea as an EPDTA. This model is quite adaptable for other species of fish or soles of different environments by varying the different characteristic functions and probability tables that are embedded in the model itself. The quantitative information (lengths, probabilities, offspring estimation) was elaborated in collaboration with the Institute of Marine Sciences of Ancona, Italy and members of their project \textbf{SoleMon} \cite{Solemon,Fabi2007}. As usually for this kind of fish, sole are categorized into so called \textit{classes} which represents soles of similar age/length and thus with similar behaviour and subject to similar natural mortality or fishing. In the Adriatic sea, according to the project \textit{SoleMon}, sole of age class 0+ aggregates inshore along the Italian coast, mostly in the area close to the Po river mouth; age class 1+ gradually migrates off-shore and adults concentrate in the deepest waters located at South West from Istria peninsula. Growth analyses on this species have been made using otoliths, scales and tagging experiments.
An otolith is a structure in the saccule or utricle of the inner ear, specifically in the vestibular labyrinth \cite{Lagardere1997}, whose section presents several concentric rings, very much like those of the tree trunks. By measuring the thickness of individual rings, it has been assumed (at least in some species) to estimate fish growth because fish growth is directly proportional to otolith growth. However, some studies disprove a direct link between body growth and otolith growth.

A great variability in the growth rate was noted: some specimens had grown 2 cm in one month, while others, of the same age group, needed a whole year. The von Bertalanffy growth function \cite{Bertalanffy1938} (VBGF) introduced by von Bertalanffy in 1938 predicts the length of a fish as a function of its age:
$$
f_{VB}(age) = L_{\infty} \left [ 1 - e^{-K(age-t_0)} \right ]
$$
The length ($f_{VB}(age)$) obtained is expressed in centimetres while $age$ and $t_0$ are in months; the different parameters that occur in the function are partly constants and partly calculated for our specific sole case study. $L_\infty$ is not the maximum length of the animal but the asymptote for the model of average length-at-age , $K$ is the so-called \textit{Brody growth rate coefficient} which, if varied, allow to manipulate the growing function in order to represent periods of low food or abundance of food (so the soles grow less or more having the same age) and $t_0$ is the time or age when the average size is zero. There parameters of VBGF for the Adriatic sole have been calculated using various methods. Within the framework of the SoleMon project, growth parameters of sole were estimated through the length-frequency distributions obtained from surveys. The results are $L_\infty = 39,6 \, cm$, $K= 0.44$ and $t_0 = -0.46$. With this correspondence we can calculate the length of a sole of age $age$ and consequently put it in one of the length classes. In the following table the ranges of the classes are shown:

\begin{center}
\begin{tabular}{|c|r|r|}
\hline
\textbf{Class} & \textbf{Minimum length} (cm) & \textbf{Maximum length} (cm) \\
\hline
0 & 0 & 18.3\\
\hline
1 & 18.4 & 25.8\\
\hline
2 & 25.9 & 30.7\\
\hline
3 & 30.8 & 33.9\\
\hline
4+ & 34 & 39.6\\
\hline
\end{tabular}
\end{center}

Considering the relevance of $K$ for the purpose of the growth function we defined in general a function $f_{VB}(age,t)$ where the parameter $t$ is an absolute month such that different periods could have a different $K$. The absolute month $t=0$ can be linked to a particular month of a particular year: in this way known past periods of low food or other environmental events can be represented in the growing function of the model. In the simulation of Section~\ref{sec-results} we used always the same $K$ along time.
Knowing the length, it is possible to estimate the weight using the length-weight relationship:\\
$ Weight(l) = a \cdot l^b $. The parameters have been estimated in SoleMon: $a = 0.007$, $b=3.0638$. Using this relationship we can determinate, at every instant during our simulations, the biomass of the whole stock under simulation.

Natural mortality (not including fishing) has been estimated through a mortality index $M$ available from the SoleMon project. From this annual index a probability distribution has been derived: $\Pr_M(i,t)$ is the probability that in a given month $t$ a sole in class $i$ dies for natural mortality. Fixing a specific month for $t=0$ the values of the function are cyclic of a period of 12 months. However, in a simulation of several years the index can be varied in different years and months with a very fine granularity. This permits to represent in a simulation catastrophic periods or particularly favourable ones. The mortality probabilities $\Pr_M(i,t)$ (on a month basis per class) used in the simulations whose results are shown in Section~\ref{sec-results} are the following:

{\small
\begin{center}
\begin{tabular}{|c|c|c|c|c|c|c|c|c|c|c|c|c|}
\hline
\textbf{Class} & \textbf{Jan.} & \textbf{Feb.} & \textbf{Mar.} & \textbf{Apr.} & \textbf{May} & \textbf{Jun.} & \textbf{Jul.} & \textbf{Aug.} & \textbf{Sep.} & \textbf{Oct.} & \textbf{Nov.} & \textbf{Dec.} \\
\hline
0 & 0.083 & 0.078 & 0.073 & 0.068 & 0.063 & 0.058 & 0.058 & 0.053 & 0.048 & 0.043 & 0.038 &  0.033 \\
\hline
1 & 0.032 & 0.031 & 0.030 & 0.023 & 0.030 & 0.028 & 0.028 & 0.028 & 0.027 & 0.026 & 0.026 & 0.025 \\
\hline
2 & 0.024 & 0.024 & 0.023 & 0.023 & 0.023 & 0.023 & 0.023 & 0.022 & 0.022 & 0.022 & 0.021 & 0.021 \\
\hline
3 & 0.021 & 0.021 & 0.021 & 0.021 & 0.021 & 0.021 & 0.021 & 0.021 & 0.021 & 0.021 & 0.021 & 0.021 \\
\hline
4+ & 0.020 & 0.020 & 0.020 & 0.020 & 0.019 & 0.019 & 0.019 & 0.019 & 0.019 & 0.019 & 0.018 & 0.018 \\
\hline
\end{tabular}
\end{center}
}

Mortality for fishing is estimated by a fishing index $F$. With the same reasoning done for the natural mortality a probability has been derived: $\Pr_F(i,t)$ is the probability that in a given month $t$ a sole in class $i$ is fished. In this case the periods of no fishing can be represented in the model. Similarly to the previous case the probability table can be cyclic over years or can be personalised month per month. The fishing index can be $F=0$, meaning that there is no fish, can be moderate (estimated $F=0.2$) or can be so strong that a situation of overfishing may occur (typically $F>1$). For instance, the fishing probabilities $\Pr_F(i,t)$ (on a month basis per class) corresponding to a fishing index $F=0.2$ are the following:

\begin{center}
\begin{tabular}{|c|c|c|c|c|c|c|c|c|c|c|c|c|}
\hline
\textbf{Class} & \textbf{Jan.} & \textbf{Feb.} & \textbf{Mar.} & \textbf{Apr.} & \textbf{May} & \textbf{Jun.} & \textbf{Jul.} & \textbf{Aug.} & \textbf{Sep.} & \textbf{Oct.} & \textbf{Nov.} & \textbf{Dec.} \\
\hline
0 & 1.65 & 1.65 & 1.65 & 1.65 & 1.65 & 1.65 & 1.65 & 1.65 & 1.65 & 1.65 & 1.65 & 1.65 \\
\hline
1 & 1.65 & 1.65 & 1.65 & 1.65 & 1.65 & 1.65 & 1.65 & 1.65 & 1.65 & 1.65 & 1.65 & 1.65 \\
\hline
2 & 1.65 & 1.65 & 1.65 & 1.65 & 1.65 & 1.65 & 1.65 & 1.65 & 1.65 & 1.65 & 1.65 & 1.65 \\
\hline
3 & 1.65 & 1.65 & 1.65 & 1.65 & 1.65 & 1.65 & 1.65 & 1.65 & 1.65 & 1.65 & 1.65 & 1.65 \\
\hline
4+ & 1.65 & 1.65 & 1.65 & 1.65 & 1.65 & 1.65 & 1.65 & 1.65 & 1.65 & 1.65 & 1.65 & 1.65 \\
\hline
\end{tabular}
\end{center}

Breeding is another important aspect of the life of soles that has been embedded in the model. In this case two estimations are needed. The first one is the probability of being reproductive after $m$ months since the last breed, that we denote $\Pr_R(m)$. For simplicity, this probability has been estimated as 0 for $m=0,1,\ldots, 11$ and as $1$ for all $m \geq 12$. However, this can be changed and refined in future versions of the model. If a sole passes this check of fertility then there is the probability of breeding $\Pr_B(i,t)$. In this case, of course, soles in class 0 have this probability equals to 0. Soles of higher classes have higher probability to breed, but only in the appropriated months, which are from November to March of every year. This probability is then spread along these months. The table used in our simulations is the following:

\begin{center}
\begin{tabular}{|c|c|c|c|c|c|c|c|c|c|c|c|c|}
\hline
\textbf{Class} & \textbf{Jan.} & \textbf{Feb.} & \textbf{Mar.} & \textbf{Apr.} & \textbf{May} & \textbf{Jun.} & \textbf{Jul.} & \textbf{Aug.} & \textbf{Sep.} & \textbf{Oct.} & \textbf{Nov.} & \textbf{Dec.} \\
\hline
0 & 0.3 & 0.25 & 0.1 & 0 & 0 & 0 & 0 & 0 & 0 & 0 & 0.1 & 0.25 \\
\hline
1 & 0.3 & 0.25 & 0.1 & 0 & 0 & 0 & 0 & 0 & 0 & 0 & 0.1 & 0.25 \\
\hline
2 & 0.3 & 0.25 & 0.1 & 0 & 0 & 0 & 0 & 0 & 0 & 0 & 0.1 & 0.25 \\
\hline
3 & 0.3 & 0.25 & 0.1 & 0 & 0 & 0 & 0 & 0 & 0 & 0 & 0.1 & 0.25 \\
\hline
4+ & 0.3 & 0.25 & 0.1 & 0 & 0 & 0 & 0 & 0 & 0 & 0 & 0.1 & 0.25 \\
\hline
\end{tabular}
\end{center}

\begin{figure}
  \begin{center}
    \includegraphics[width=14cm]{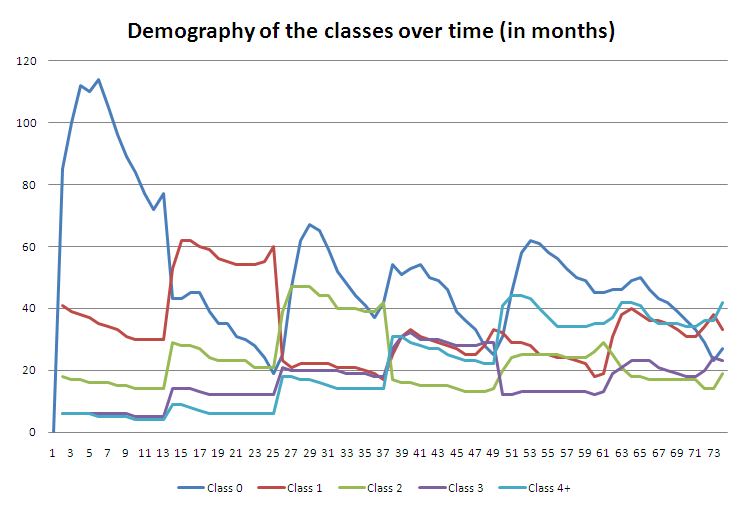}
  \end{center}
\caption{Demography over time without fishing ($F=0.0$).}
\label{demography0}
\end{figure}

\begin{figure}
  \begin{center}
    \includegraphics[width=14cm]{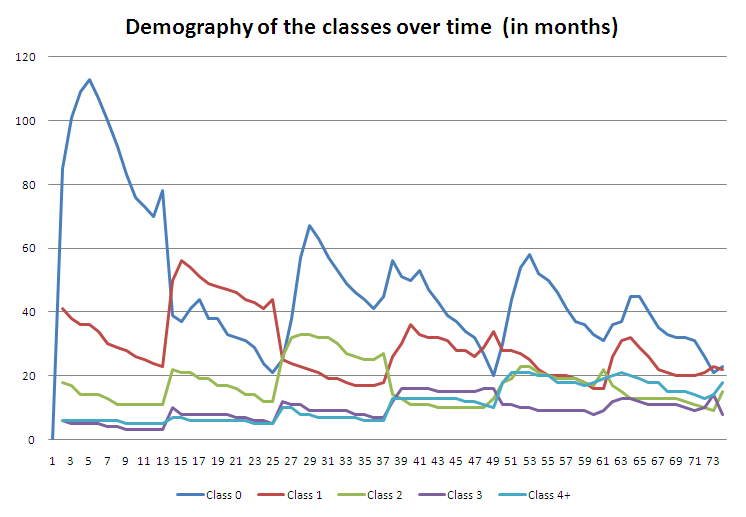}
  \end{center}
\caption{Demography over time with light fishing ($F=0.2$).}
\label{demography02}
\end{figure}

\begin{figure}
  \begin{center}
    \includegraphics[width=14cm]{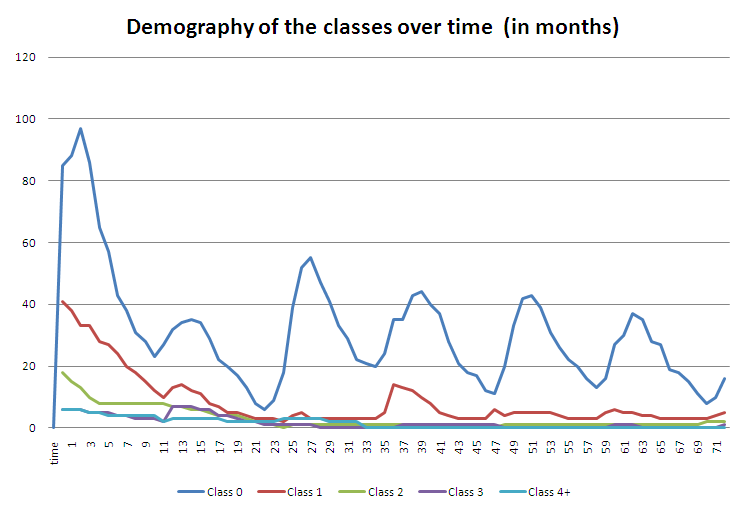}
  \end{center}
\caption{Demography over time with overfishing ($F=1.2$).}
\label{demography12}
\end{figure}

Note that in the case of breeding there is a potential artificial situation in a simulation. It can happen that one sole of the higher classes do not breed at all along one year. It is unlikely, but in the simulation may happen. This is absolutely not possible in reality. We plan to adapt the model in the future in order to avoid this situation. Another weakness of the model (and obviously of the simulator) is the lack of information about the offspring of an individual. This is a forced missing because no real information is known about the number of eggs that are fecundated nor the number of eggs that hatch, becoming new individuals. This issue is managed from the tool for the purposes of the simulation and will be treated in details in Section \ref{sec-fishpass}.

The resulting overall model is shown in Figure~\ref{fig-solea}. Two clocks are used, $x$ for counting the passage of one month and $t$, never reset, for measuring the absolute time since the beginning. Note that only the transitions from one generic location ``class $i$'' are shown. The whole model is simply the resulting EPDTA considering the locations for all the classes, while the locations ``dead'' and ``fished'' are the same for all the classes.
Every class has its particular fishing, mortality and breeding probability (e.g.\ a smaller, thus younger sole, has less probability to die/be fished than a older one). The boolean variables $M_c, F_c, R_c$ are used to assure that every month the sole makes a mortality check ($M_c$), a fishing check ($F_c$) and a reproduction/breeding check ($R_c$). Note that time can advance of one month only if all these checks are done. The urgency\footnote{In the picture the urgent transitions are indicated by a little ``u'' attached at the beginning of their arrows.} is used for forcing the class change of the class as soon as the sole reaches the minimum length for that class.
The labels $\mathrm{dead}\_i$, $\mathrm{fish}\_i$ and $\mathrm{breed}\_i$ are used to communicate to the simulation environment that a particular sole (the one sending the signal) of class $i$ died, was fished or breed. The meaning of the integer variables $age$, $length$ is obvious and the variable $lastB$ counts the months elapsed since the last breeding of this sole. The probability tables and the functions used in Figure have been all described above.

\section{Simulation}
\label{sec:simul}
As shown at the end of Section~\ref{sec:epdta} the probability model can be automatically checked to discover interesting properties or to make consistency checks on the given probabilities. Another important way to use the same model is for simulating a sole in a population of them. The idea is not new, but it very naturally fits in the fish stock monitoring. If we want to predict what happens to a population after several years of fishing of a certain strength under normal or particular conditions, all we have to do is to instruct a population of virtual soles that uses the given model as probabilistic behaviour and let them evolve over time. In every moment, our virtual environment can show to us all the statistics we want to know about the whole stock but also about every single sole.

Naturally the hard thing to do is to precisely tune the model with the most possible available real data. This work has to be done before the simulations on the model can be considered to have a certain degree of reliability.

In this section we show the simulator that we have developed for reaching this goal. It uses agent technology, as we discuss in the following. We are currently at the very initial phase of the tuning of probabilities and other values using real data. The implementation is quite stable and the adaptability to other species can be done quite easily. More information is available at \cite{FishpassSite}.

\subsection{Multi-Agent Systems}
Several definitions have been given for the term ``agent'' during the last decades, the most suited of which is the one given from Russel: an agent is something that can retrieve information from the environment through its sensor and can perform actions with its actuator \cite{Russell2003}. Alternatively, Woldrige and Jenning \cite{Jennings1998,Wooldridge1994} define agent as hardware or software-based computer system that have the following properties: autonomy, reactivity, pro-activeness, and social ability. A \textbf{Multi-Agent System} (MAS) is a collection of autonomous agents that communicate, cooperate, share knowledge and solve their own problem.

In a MAS, each agent can be either cooperative or selfish; in other words the single agent can share a common goal with the others (e.g. an ant colony), or they can pursue their own interests (as in the free market economy). MAS are usually exploited when the problem considered cannot be solve (efficiently) by an individual agent or a monolithic system. They are used to model coordinated defence systems but also for disaster response models, social network modelling, transportation, logistics, graphics as well as in many other fields when the problem is non-linear or the interaction with flexible individual participants have to be represented or again when in-homogeneous space is relevant. Finally, MAS are widely used in networking and mobile technologies, to achieve automatic and dynamic load balancing, high scalability, and self-healing networks.

In the context of a MAS, an agent needs to \textit{communicate} its information to the others and after that it needs to \textit{coordinate} its activities (which is important to prevent conflicts between the agent belonging to the MAS) and \textit{negotiate} its interest to solve a problem without conflicts. This need of interaction and exchange of information between agents is the basic characteristic that differentiate MASs from traditional artificial intelligence which work only as a single agent.

\subsection{Hermes middleware} 
Hermes \cite{Corradini2005,HermesSite} is an agent-based middleware for designing and execution of activity-based applications in distributed environments. It provides an integrated environment where users can focus on the design of the particular activity of interest ignoring the topological structure of the distributed environment. Hermes consists of a 3-layer software architecture: the Agent layer, the BasicServices layer and the Core layer.

An Hermes execution consists of a creation of a MAS in which the agents are of two kinds: user agents and service agents. The Agent Layer is the upper layer of the mobile platform that contains both kinds of agents. A service agent accesses to local place resources such as data and tools (which, for security reason, are not directly accessible) while a user agent executes complex tasks and implement part of the logic of the application. Hermes is based on the concept of \textit{place}: a place is a well defined node of a network where service agents are located. When a service agent is created on a place and bound to it, there is no way for it to migrate to another place of the network. User agents can instead be copied to another place (weak mobility) and their execution can continue on the migration place.

\subsection{FIShPASs: FIshing Stock Probabilistic Agent-based Simulator}
\label{sec-fishpass}

FIShPASs \cite{FishpassSite} is a simulator based on Hermes. It exploits the agent paradigm to simulate, as a MAS, the evolution of a population of fish of a certain species. At the moment a model for a sole (Solea solea) population living in the North Adriatic sea is available, but the simulator can be easily adapted for other species or soles of different environments

The basic elements of the simulator are: the \texttt{SoleaAgent}, the \texttt{SquareKilometreSea}, the \texttt{Registry} and the \texttt{Randomizer}. The \texttt{SquareKilometreSea} is exactly the Hermes place where the simulation is started. This first version of the simulator is quite simple since the sea is not spatially simulated; it is more like a \textit{simple container} for the population of soles (exactly what is an Hermes place for its agents). In the near future we point to improve the overall model with support to space and thus displacement of sole in the simulated sea, water currents, temperature and so on.

The \texttt{SoleaAgent} is a user agent and represents a single sole in the simulated sea. While the spatial model is not so accurate, the sole \textit{behavioural} model is quite complex. Indeed the \texttt{SoleaAgent} implements the EPDTA presented in Section~\ref{sec-model} and thus can be fished, can die for natural mortality, can reproduce with the given probabilities, and naturally grows as time passes. Note that also the non-deterministic choices that are presented to the agent (as its behaviour is essentially a Markov decision process) are resolved probabilistically with a uniform distribution on all the enabled non-deterministic choices. As briefly discussed in Section~\ref{sec-model}, since the simulation is managed on a month basis, we can arrange theoretical probability so that certain behaviour cannot occur or can occur rarely in certain periods of the year. For instance, we can suppose that the fishing period goes (hypothetically) from October to February while in the other months fishing is prohibited and fix the fishing probability to 0 for the prohibited period. Since the probability values are given also on the basis of the length of the sole, the model can be easily adapted to different scenarios to simulate, for example, overfishing of some classes or sudden reduction of the population of some other classes because of a disease and so on.

The third element of the simulation is the \texttt{Randomizer}. It is a service agent which generates random numbers in $[0,1]$  for the sole. Every time a sole checks the probability of doing something (death, reproduction, etc.) it requests a new number to the \texttt{Randomizer}. The service returns a new generated value that is contrasted with the probability of the individual to decide whether the action occurs or not.

The last main element of the simulator is the \texttt{Registry}. Like the \texttt{Randomizer} it is a service agent and it simply keeps track of the sole available in the simulated sea. Its main purposes are the consistency control over the simulation and the generation of statistics about the simulated months (in particular, population per different class, weight of the biomass, number of death/fished soles in the last month). At the beginning of the simulation the \texttt{Registry} reads the input data and computes the number of individuals of the initial population then waits for them to communicate their status to the registry itself.

\texttt{SoleAgent}s are programmed to communicate their status to the \texttt{Registry} every month in any case (even in case of death), which means that the \texttt{Registry} can always know if all the soles have communicated with it during the current interaction. Moreover, it always knows the exactly demography of the population. This communication acts also as a synchronization that ensure time consistency on the population. It is the \texttt{Registry} that manages time increments and that enables the sole to execute their internal behaviour (setting the \texttt{SoleaAgent} variable corresponding to the local clock ``x'' of the model of Figure~\ref{fig-solea} to 1). Before each increment the \texttt{Registry} waits for a communication from all the population of soles and then it increments the time. In this way the \texttt{Registry} ensure that at each simulation step (i.e. each month) no sole is out of simulation time range (behind or beyond the current time).

Finally, the \texttt{Registry} generates new soles if some of the existing ones reproduced during the last month. Having the generation in the \texttt{Registry} is a strategic choice to be sure that the new born sole will be correctly set in the current time frame. In reality, a female sole produces, depending on its class, between 150000 and 250000 eggs and spreads them in the water. The number of them that will grow at least until class 1 is very low. Currently there is not a direct known relation between the number of females that breed in a month or in a season and the number of surviving and developing eggs in the following months/season. Thus, the \texttt{Registry} creates every month a number of newborn soles in class 0 that corresponds to the number observed in reality (data from SoleMon project). One challenge for the future will be trying to find a relation between the signals of breed ($\mathrm{breed}_i$) given by the \texttt{SoleAgent}s to the \texttt{Registry} in a certain period and the number of surviving newborn to introduce after some month(s).

Given the real data about individuals in the different classes from 2005 to 2008, we taken the first column, which represents the newborns (males + females) of every year, halved the values (since we consider only female soles) and distributed the newborns so obtained along the year, according to the previous fertility table.

\begin{center}
\begin{tabular}{|c|c|c|c|c|c|c|c|c|c|c|}
\hline
\textbf{population km$^2$} & \textbf{0} & \textbf{1} & \textbf{2} & \textbf{3} & \textbf{4+} \\
\hline
2005 & 169 & 82 & 36 & 12 & 4  \\
\hline
2006 & 92 & 179 & 43 & 10 & 1  \\
\hline
2007 & 205 & 138 & 72 & 18 & 1 \\
\hline
2008 & 117 & 123 & 61 & 10 & 6 \\
\hline
\end{tabular}
\end{center}

In such a way we obtain the \textit{birth rate} table below. It represents the amount of newborns that are \textit{automatically} generated from the simulator every simulated month. Since the table covers only 4 years it is used cyclically in the subsequent years, thus the fifth year the generated newborns come from the first row of the table and so on.

\vspace{0.1cm}

\begin{center}
\begin{tabular}{|c|c|c|c|c|c|c|c|c|c|c|c|c|}
\hline
\textbf{Year} & \textbf{Jan.} & \textbf{Feb.} & \textbf{Mar.} & \textbf{Apr.} & \textbf{May} & \textbf{Jun.} & \textbf{Jul.} & \textbf{Aug.} & \textbf{Sep.} & \textbf{Oct.} & \textbf{Nov.} & \textbf{Dec.} \\
\hline
1 & 26 & 21 & 9 & 0 & 0 & 0 & 0 & 0 & 0 & 0 & 8 & 21 \\
\hline
2 & 14 & 12 & 4 & 0 & 0 & 0 & 0 & 0 & 0 & 0 & 4 & 12 \\
\hline
3 & 30 & 25 & 11 & 0 & 0 & 0 & 0 & 0 & 0 & 0 & 11 & 25 \\
\hline
4 & 16 & 15 & 6 & 0 & 0 & 0 & 0 & 0 & 0 & 0 & 6 & 15 \\
\hline
\end{tabular}
\end{center}

\vspace{0.1cm}

Summing up, the FIShPASs simulation steps are the following:
\begin{enumerate}
  \item the sea (place) is launched and the service agents (\texttt{Randomizer} and \texttt{Registry}) are generated on it. The \texttt{Registry} calculates the initial population
  \item the sole population is generated from the place basing on the SoleMon project data
  \item the soles register to the \texttt{Registry}
  \item once all population has signed to the \texttt{Registry}, it generates statistics and starts their behavioural simulation by sending them a message to update their internal clock $x$
  \item the soles execute all their operations for the current month, reset the clock $x$ and send an ack to the \texttt{Registry}
  \item once all the acks are received by the \texttt{Registry}, it generates statistics for the elapsed month, creates newborn soles and then sends a new message to increment the clock $x$
\end{enumerate}
The last two points are repeated until the desired period of time has passed.

\begin{figure}
  \begin{center}
    \includegraphics[width=14cm]{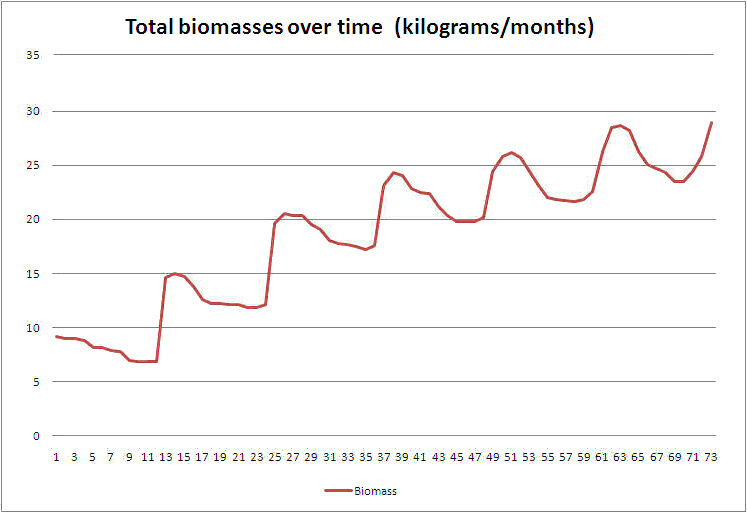}
  \end{center}
\caption{Biomass over time without fishing ($F=0.0$.)}
\label{Biomasses0}
\end{figure}

\begin{figure}
  \begin{center}
    \includegraphics[width=14cm]{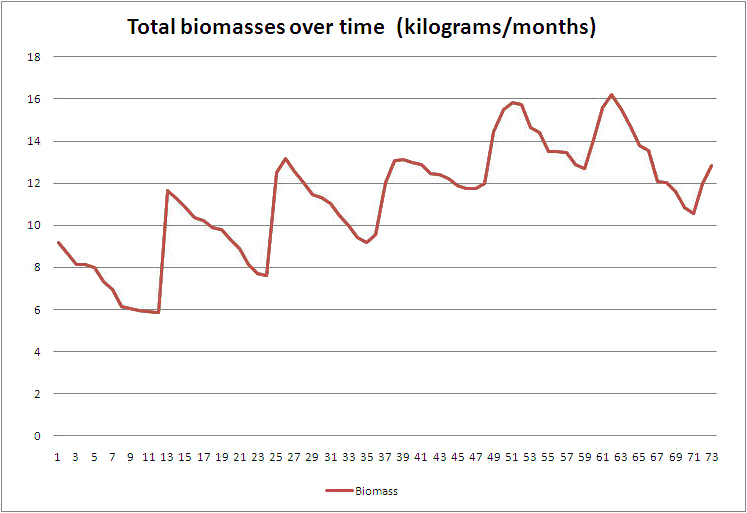}
  \end{center}
\caption{Biomass over time with overfishing ($F=1.2$).}
\label{Biomasses02}
\end{figure}

\begin{figure}
  \begin{center}
    \includegraphics[width=14cm]{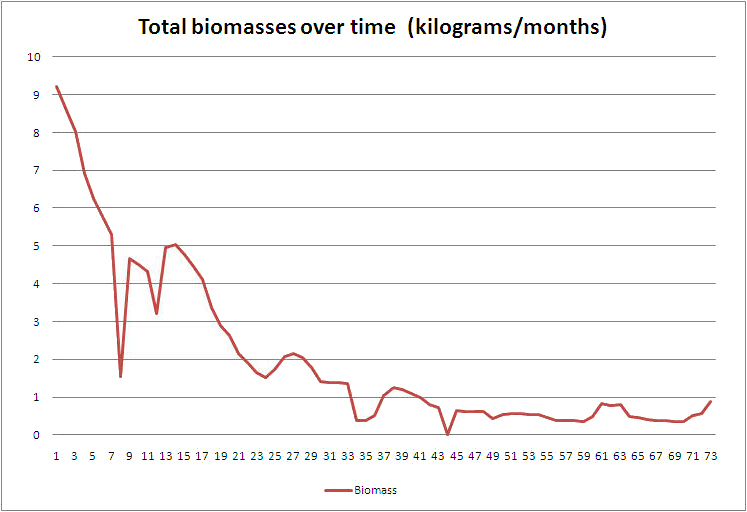}
  \end{center}
\caption{Biomass over time with overfishing ($F=1.2$).}
\label{Biomasses12}
\end{figure}

\subsection{Results}
\label{sec-results}

We set up a series of simulations to test our model. Every simulation has a timestep one month long and lasts for 72 months (6 years) considering always a virtual square kilometre of sea. The three charts of Figures~\ref{demography0},~\ref{demography02}~and~\ref{demography12} show the trend of population that can occur, along the simulation, varying the fishing probability (i.e.\ the probability for a sole to be fished) while mortality and reproduction probabilities remain the same. The charts concentrate on three very different scenarios. The first one considers a case of no fishing (fishing index F=0.0). In this case the population remains stable, spreading on the different classes. It is particularly notable an increment of soles in the third class as well as a gradual and steady increment of individual in the fourth class.

The second case considers a light fishing activity (F=0.2). The scenario rapidly changes with classes third and fourth that grow slower than in the previous chart. Moreover all of them has difficulty to get over 20 individuals (whereas in the previous plot all of the bigger classes where around 40 individuals). The trend is that of a decreased population of soles composed of less mature individuals and some small ones (see the biomasses charts for more details).

The third and last chart about the population shows another possible scenario. In this case we suppose that the sea undergoes an overfishing activity (F= 1.2). This situation has obviously an extreme impact on the population. As can be seen, after two years (23-24 on the x-axis), the population is simply gone with only the 0 class of individuals available (due to their automatic generation, as explained above).

The three charts of Figures~\ref{Biomasses0},~\ref{Biomasses02}~and~\ref{Biomasses12} represent the biomass trend, i.e.\ the total amount (in kilograms) of soles for the different classes in the three scenarios described above. In the case of no fishing (F=0.0) the sole population tends to spread over all the classes and the biomass grows accordingly. The biomass increment is constant and at the end of the 6 years the total biomass is around 28 kg (against 8 kg at the beginning of the simulation).

When we introduce light fishing (F=0.2) the biomass tendency is similar to the previous scenario but the values are totally different. In particular the population is impoverished in the higher classes (see previous plots) and the overall biomass grows slower at the beginning with a more marked decreasing tendency during the last year (months 60 - 73) when soles grow, reaching class fourth and thus are easier to be fished. At the end of the simulation the total biomass is around 13 kg which means less than an half of the soles biomass without fishing.

With the introduction of overfishing (F=1.2) the scenario changes drastically. The fishing activities has a great impact on the population biomass that is halved at the beginning of the second year (13-15 on the x-axis) and then runs fast under 1 kg in the middle of the forth year (41-42 on the x-axis). All the bigger soles, more subject to fishing, have been caught or are dead and only the smaller ones (i.e. with a small biomass) remain (also because they are auto-generated). Again, as seen in the corresponding classes chart, the population is decimated.

More details and charts about these simulations can be found on the simulator website \cite{FishpassSite} along with contacts to request a copy of the current version of the tool.

\section{Conclusions and Future Work}
\label{sec:concl}
We have defined an individual-based model of the behaviour of a common sole (\textit{Solea solea}) living in North Adriatic sea. The model has been specified as an Extended Probabilistic Discrete Timed Automaton (EPDTA), a formalism that is a variant of probabilistic timed automata. We have defined the semantics of an EPDTA as a Markov decision process and we have observed that an EPDTA can be translated to a syntax acceptable by the model-checker PRISM. The estimation of the probabilities and of the characteristic function of the species has been done by using the real data of the SoleMon project. The individual probabilistic behaviour then has been embedded into an agent of a MAS. The MAS simulates the population of soles over time and can provide information on the evolution of the stock by monthly statistics of the individual states. We have presented the simulator FIShPASs (FIshing Stock Probabilistic Agent-based Simulator) that implements the presented model and is easily adaptable for other species.

There are a lot of interesting things to do as future work. First, we want to tune the model,  working in team with specialized biologists, in order to increase the confidence on its predictions. The translation of the model into a PRISM acceptable syntax can be made available inside the simulation environment. Having the PRISM equivalent model can highly improve the tests that the biologists can do against the probabilities put in the various tables embedded in the model. This is because quantitative questions can be asked to the model checker to test hypothesis made about the model itself or to validate it with available real data. In the MAS part a huge number of improvements are possible. For instance, soles can be given a geometrical space to occupy and can move in the simulated square kilometre. They can also emigrate and immigrate from/to the simulated space. A 3D environment, i.e.\ a cube kilometre, instead of a 2D one could be more appropriate because other species could be simulated simultaneously and made interact with the soles (towards a more predator-prey approach). Moreover, a physical conformation of the territory can be added to the model possibly influencing the interactions (of different kind, to be introduced in the model too) between the individuals (the formation of an isolated population, the impossibility to meet, etc.). Finally, the effects of the passage of a particular fishing device can be modelled; for this we know there are available data for tuning/validation.

\bibliographystyle{eptcs} 
\bibliography{soleanew}
\end{document}